\documentclass[useAMS,usenatbib]{mn2e}
\usepackage{graphicx}
\def\WD{{\it WD}}
\def\WO{{\it WO}}
\def\apj{ApJ}
\def\aap{A\&A}
\def\iaucirc{IAU Circ.}
\title[V4633 Sgr -- a probable second asynchronous polar classical nova]
{V4633 Sgr -- a probable second asynchronous polar classical nova}
\author[Y.~M.~Lipkin and E.~M.~Leibowitz]
{Y.~M.~Lipkin$^1$\thanks{E-mail: yiftah@wise.tau.ac.il (YML);
    elia@wise.tau.ac.il (EML)} and 
    E.~M.~Leibowitz$^1$\footnotemark[1]\\
 $^1$School of Physics and Astronomy and the Wise Observatory, Raymond
 and Beverly Sackler Faculty of Exact Sciences,\\ Tel-Aviv University,
 Tel Aviv, 69978, Israel}

\begin{document}

\date{Accepted 2008 mmmmmm dd. Received 2008 February 07; in original form 2008 January 28}

\pagerange{\pageref{firstpage}--\pageref{lastpage}} \pubyear{2008}

\maketitle

\label{firstpage}

\begin{abstract}
Photometric observations of V4633~Sgr (Nova Sagittarii 1998) during
1998-2005 reveal the presence of a stable photometric periodicity at 
$P_1=180.8$~min which is probably the orbital period of the underlying
binary system.
A second period was present in the light curve of the object for six
years.
Shortly after the nova eruption it was measured as $P_2 = 185.6$~min.
It has decreased monotonically in the following few years reaching the
value $P_2 = 183.9$~min in 2003.
In 2004 it was no longer detectable.
We suggest that the second periodicity is the spin of the magnetic
white dwarf of this system that rotates nearly synchronously with the
orbital revolution.
According to our interpretation, the post-eruption evolution of Nova
V4633 Sgr  follows a track similar to the one taken by V1500~Cyg (Nova
Cygni 1975) after that nova eruption, on a somewhat longer time
scale.
The asynchronism is probably the result of the nova outburst that lead
to a considerable expansion of the white dwarf's photosphere.
The increase in the moment of inertia of the star was associated with
a corresponding decrease in its spin rate.
Our observations have followed the spinning up of the white dwarf 
resulting from the contraction of its outer envelope as the
star is slowly retuning to its pre-outburst state.
It is thus the second known asynchronous polar classical nova.
\end{abstract}

\begin{keywords}
binaries: close -- stars: individual: V4633~Sgr -- novae, cataclysmic variables.
\end{keywords}

\section{Introduction \label{Sec:V4633Sgr:Introduction}}
Nova Sagittarii 1998 was discovered by \citet{Liller1998} on 1998 March
22.4, and reached maximum brightness of 7.4 mag on March 23.7 (Jones
1998).
The nova was a fast one, with $t_3\approx35$~d in visual and
$\approx49$~d in $V$ \citep{LillerJones1999}.

\defcitealias{LipkinEtAl2001}{L01}
In an earlier paper (\citealt{LipkinEtAl2001}, hereafter
\citetalias{LipkinEtAl2001}) we reported on the presence of two
photometric signals, early after the eruption: a coherent $P_1$ =
3.014~h periodicity (hereafter the primary signal), and an unstable
$P_2 \approx $3.09~h signal that decreased by 0.3\% during three
years of observations (hereafter the secondary signal).
We suggested that the coherent primary signal is the orbital
periodicity of the underlying binary system and that the secondary
signal is either superhumps or the spin period of a spinning-up
magnetic {\WD}.
In the latter interpretation, V4633~Sgr is an aynchronous polar
(hereafter, AP) -- a strongly-magnetic CV in which the primary's spin
period marginally differs from the orbital period (by $\sim1$\%, see,
e.g., \citealp*{Ref:V4633Sgr:2003A_A...407..987Staubert}) , and the {\WD}
spin-up is caused by the post-eruption contraction of the
strongly-coupled, expanded  {\WD} envelope (see
\citealp*{Ref:V4633Sgr:Stockman_Schmidt_Lamb1988} for a description of
the spin dynamics in the prototype, V1500~Cyg).

Preliminary analysis of follow-up observations in 2001-2002 revealed
the continued decrease of $P_2$, which measured 3.07~h in 2002
\citep{Lipkin2004}.
This further strengthned the asynchronous polar interpretation for the
system.

Here we report on the results of observations of two additional years,
and on the analysis of our whole accumulated data-set, including a
reanalysis of the published data.

\section{The long-term light curve\label{Sec:V4633Sgr:LC}}
\begin{figure}
\includegraphics[width=84mm]{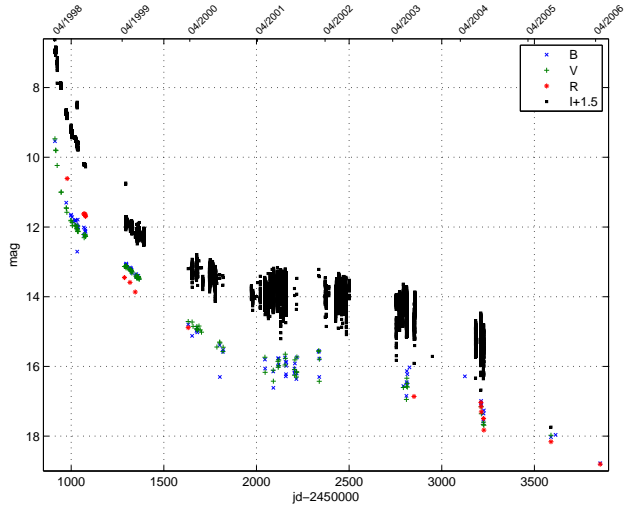}
\vskip 0.07in
\caption{The $BVRI$ LCs of V4633~Sgr in 1998-2006. The $BVR$
  plots are nightly averages of our observations, while the $I$ plot
  is our full data. The $I$ LC is shifted up by
  1.5~mag for better viewing.\label{Fig:V4633Sgr:LC}}
\end{figure}

The $BVRI$ light curves (LC) of v4633~Sgr accumulated at the Wise
Observatory {\WO} during the first 8~yr following its eruption are
shown in Fig.~\ref{Fig:V4633Sgr:LC}.
These comprise some 16500 images accumulated in 238 nights during
1998-2006.
Table \ref{Table:V4633Sgr:IObs} summarizes our $I$-band observations.
After the eruption, a 3-yr phase of brightness-decline followed,
during which the nova has faded by approximately 8~mag.
This initial brightness decline phase ceased in 2001, when a 
post-nova roughly stable brightness state of 15.5-16~mag has been
reached.
In 2003, a new phase of brightness-decline commenced, and the system
faded during the following 4 years by $\sim$3~mag to 18-19~mag in 2006
at the end of our observations.

\begin{table}
  \centering
  \begin{minipage}{80mm}
    \caption{A summary of the $I$-band observations of
      V4633~Sgr (note that in 2006 only a few $B$ and $R$, but no $I$  snapshot
      were observed) \label{Table:V4633Sgr:IObs}}
    \begin{tabular}{@{}l|cccccccc@{}}
      \hline
      Year   & 1998 & 1999 & 2000 & 2001 & 2002 & 2003 & 2004& 2005 \\
      \hline
      Images & 4035 & 2103 & 1205 & 2317 & 1083 & 1226 & 683 & 1\\
      Nights & 33   & 34   & 38   & 58   & 28   & 20   & 12  & 1\\
      \hline
    \end{tabular}
  \end{minipage}
\end{table}

\section{The primary signal\label{Sec:V4633Sgr:P1}}
\begin{figure}
\includegraphics[width=84mm]{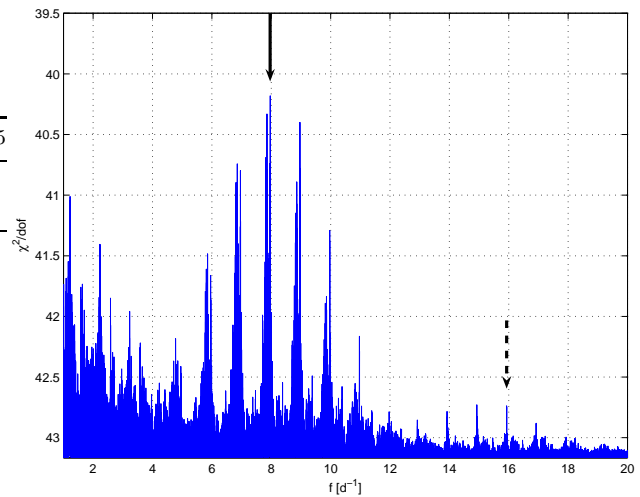}
\vskip 0.07in
\caption{A Periodogram of the 1999-2004 $I$-band data of
  V4633~Sgr. The y-axis is the $\chi^2/{\rm dof}$ obtained by fitting
  the data to a 7th-degree polynomial and a sine wave with test
  frequencies spanning the x-axis. The y-axis is flipped so that
  better-fitting (lower $\chi^2/{\rm dof}$) frequencies are peaks
  instead of troughs. The best-fitted frequency at 7.964~d$^{-1}$ is
  marked by a solid-line arrow. The second harmonic, at
  15.928~d$^{-1}$ is marked by the dashed-line
  arrow.\label{Fig:V4633Sgr:PS1}}
\end{figure}

Figure~\ref{Fig:V4633Sgr:PS1} shows a periodogram of the 1999-2004
$I$-band LC of V4633~Sgr.
The periodogram shows the $\chi^2/{\rm dof}$ values obtained by
fitting the data to a sine wave with test-frequencies spanning the
range shown in the figure, and to a 7th-degree polynomial, introduced
to detrend the post-eruption secular brightness decline of the nova.
The periodogram is dominated by the peak of the primary signal at
7.964~d$^{-1}$ (marked by a solid-line arrow in
Fig.~\ref{Fig:V4633Sgr:PS1}) and its complex alias structure,
affirming the coherence of this signal over the time spanned
by our observations.
The second harmonic of $P_1$ at twice the fundamental frequency,
15.928~d$^{-1}$, is also clearly present in the periodogram (marked by
a dashed-line arrow in Fig.~\ref{Fig:V4633Sgr:PS1}).
The best-fit value of the fundamental frequency is 
$P_1 = 180.8169\pm0.0002$~min (the quoted error is 1-$\sigma$
confidence level that was derived by a sample of 3000 bootstrap
simulations, \citealp{Efron_Tibshirani1993}).
This result is insensitive to the degree of polynomial used to
detrend the LC, for 6th- to 11th-degree polynomials.
The calculated epemeris of the primary minima is given in Equation~\ref{Eqtn:Ephem}.
\begin{equation}
\label{Eqtn:Ephem}
\rmn{HJD}\, T_{min} = 2453227.59405596 + 0.125567(27) E
\end{equation}
Measurements of $P_1$ in each observational season are also
consistent with a stable period (the yearly measurements are presented
as the $P'_1$ column in Table~\ref{Table:V4633Sgr:Periods}; the upper
limit for variation of $P_1$ is $\dot{P_1} \le 1.1\times10^{-8}$).

\begin{figure}
\includegraphics[width=84mm]{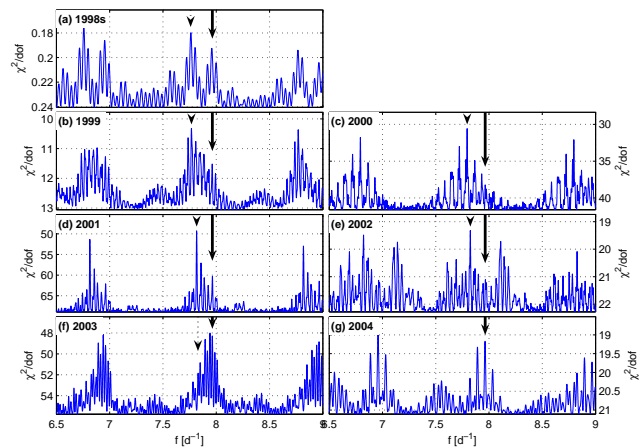}
\vskip 0.07in
\caption{Yearly periodograms of V4633~Sgr, obtained by fitting each
  yearly data to a 3rd-degree polynomial and a sine function with a
  test frequency. The primary signal is marked in each plot by a
  solid-line arrow, and the secondary signal is marked by a
  dashed-line arrow.\label{Fig:V4633Sgr:PSyr}}
\end{figure}

Periodograms of each observational season are shown in
Fig.~\ref{Fig:V4633Sgr:PSyr}.
Each data-set was fitted to a sine function and a 3rd-degree polynomial.
The primary signal is clearly present in each of the periodograms
(marked by a solid-line arrow), and is the prominent signal in the
periodograms of 2003 and 2004.

\begin{figure}
\includegraphics[width=84mm]{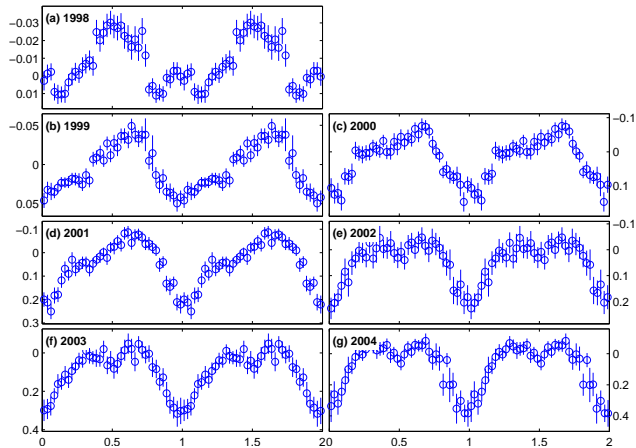}
\vskip 0.07in
\caption{The $I$ LCs of each of the observational seasons folded on
  the period of $P_1$. Discrete points are mean magnitudes in each of
  40 equal bins covering the 0--1 phase interval. The inserted bars
  are the standard deviation in the value of the mean in each bin. In
  all but the 2004 data, the signal of $P_2$ and a
  3rd-degree polynomial were subtracted prior to folding. Note the
  difference in the y-axis scales. \label{Fig:V4633Sgr:FoldP1}}
\end{figure}

The yearly waveforms of $P_1$ are shown in
Fig.~\ref{Fig:V4633Sgr:FoldP1}.
In each case, the LC was detrended prior to folding by subtracting the
secondary signal and a 3rd-degree polynomial (see
Sec.~\ref{Sec:V4633Sgr:P2}).
The initial waveform of $P_1$ in 1998 was symmetric around a single
major minimum.
A clear small symmetric hump of short duration is seen at the center
of this minimum, indicating a slight brightening of the system around
the phase of lowest point.
In the following year the minimum became asymmetric around the low
point, with a slow rise and a fast decline. It then gradually turned
more symmetric, with a primary minimum and a secondary dip of 0.1233~
mag in 2004.
Note that the primary minimum of this periodicity preserved its phase
throughout the 8 years of our observations.
If it is the phase of superior conjunction of the WD, the 1998 profile
indicates that shortly after outburst, namely within a few tens of
days following the eruption, there was a particular intense activity
in the system, either on the surface of the WD or in its close
vicinity, that took place along the interbinary center line.

The full amplitude of $P_1$ in 1998-2004 was 0.040~mag, 0.085~mag,
0.20~mag, 0.30~mag, 0.26~mag, 0.36~mag, and 0.45~mag in each of the
years, respectively.
The flux variation of $P_1$ in 1999-2004 relative to that of 1998 was
0.085, 0.069, 0.068, 0.055, 0.048, and 0.028, respectively.

\begin{table}
  \centering
  \begin{minipage}{70mm}
    \caption{The best-fit values of the primary and the secondary signals
      in each observational season. The second column ({\bf $P_2$})
      presents the results for $P_2$ when $P_1 = 180.8169\pm0.0002$~min,
      the best fit value for the entire LC, was set fixed. The third and
      fourth columns ({\bf $P'_1$} and {\bf $P'_2$}) are the results for
      $P_1$ and $P_2$ when both signals were fitted simultaneously (see
      text for a full description).\label{Table:V4633Sgr:Periods}}
    \begin{tabular}{@{}lcc|cc@{}}
      \hline
      {\bf Year} & {\bf $P_2$~[min]} &  {\bf $P'_1$~[min]} & {\bf $P'_2$~[min]}\\ 
      \hline
      1998 & $185.577\pm0.038$ & $180.711\pm0.081$ & $185.604\pm0.039$ \\
      1999 & $185.394\pm0.007$ & $180.852\pm0.008$ & $185.396\pm0.006$ \\
      2000 & $184.738\pm0.005$ & $180.817\pm0.009$ & $184.739\pm0.005$ \\
      2001 & $184.239\pm0.004$ & $180.808\pm0.007$ & $184.238\pm0.004$ \\
      2002 & $184.033\pm0.009$ & $180.768\pm0.011$ & $184.045\pm0.009$ \\
      2003 & $183.946\pm0.018$ & $180.838\pm0.009$ & $183.945\pm0.014$ \\
      2004 &          -        & $180.846\pm0.027$ &         -        \\
      \hline
    \end{tabular}
  \end{minipage}
\end{table}

\section{The secondary signal}
\label{Sec:V4633Sgr:P2}
\begin{figure}
\includegraphics[width=84mm]{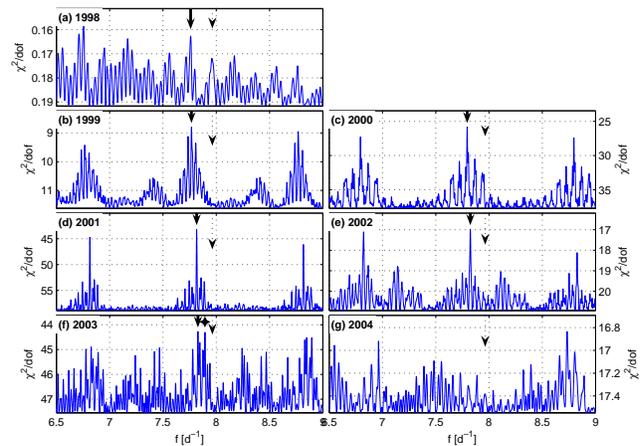}
\vskip 0.07in
\caption{The same as Fig.~\ref{Fig:V4633Sgr:PSyr}; Each yearly
  data-set was fit to a 3rd-degree polynomial, the first three
  harmonics of $P_1$, and a sine wave. In each periodogram, the
  frequency of the primary signal is marked by a dashed-line arrow,
  and the secondary signal by a solid-line arrow. The secondary signal
  is not detectable in the 2004 periodogram\label{Fig:V4633Sgr:PS1yr}}
\end{figure}

We tested the yearly light curves for a second signal by fitting each
data-set to a 3rd-degree polynomial, the first three harmonics of
$P_1$, and a sine function with test-frequencies spanning the range
5~d$^{-1}$--12~d$^{-1}$.
The resulting periodograms (hereafter, the secondary periodograms) are
shown in Fig.~\ref{Fig:V4633Sgr:PS1yr}.
The strongest signal in each periodogram, which is the secondary
signal in the LC, is marked by a solid-line arrow.
The dashed arrow marks the frequency of $P_1$.

In the primary, as well as in the secondary periodograms of the
data-sets of the first five years, 1998 -- 2002, the prominent peak
corresponds to a period $P_2$ that is slightly longer than $P_1$ (in
1998 it is the 1-day alias of what we consider the true periodicity in
the LC).
The best-fit values of $P_2$ in each yearly data-set are given in
Table~\ref{Table:V4633Sgr:Periods} (Col.~2).
The secondary period monotonically decreased at a
rate that also declined.
It was 2.6\% longer than $P_1$ a few months after the eruption,
becoming 1.8\% longer than $P_1$ four years after.
Consistent results are obtained when each yearly data-set is fit
simultaneously to $P_1$ and $P_2$ (Table~\ref{Table:V4633Sgr:Periods},
Col.~4), or by conducting the analysis using different parameters (a
lower degree polynomial, fewer harmonics of $P_1$, or more harmonics
of the test frequency).

The secondary periodogram of 2003 features a prominent signal at
$7.83$~d$^{-1}$, 1.7\% longer than the primary signal, which we
identify as the still-declining $P_2$.
A slightly lower peak at $7.89$~d$^{-1}$ (marked
with a star-headed arrow in Fig.~\ref{Fig:V4633Sgr:PS1yr}) is probably
an alias.

\begin{figure}
\includegraphics[width=84mm]{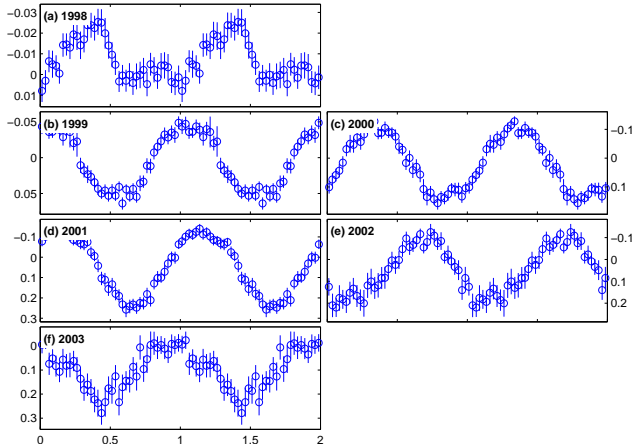}
\vskip 0.07in
\caption{The same as Fig.~\ref{Fig:V4633Sgr:FoldP1}, with the LCs folded
  on the period of $P_2$.\label{Fig:V4633Sgr:FoldP2}}
\end{figure}

The yearly waveforms of $P_2$ are shown in Fig.~\ref{Fig:V4633Sgr:FoldP2}.
The LCs were detrended prior to folding by removing $P_1$ and a
3rd-degree polynomial.
The full amplitude of $P_2$ was 0.03~mag, 0.11~mag, 0.29~mag, 0.40~mag,
0.34~mag, and 0.30~mag, in 1998-2003, respectively.

\begin{figure}
\includegraphics[width=84mm]{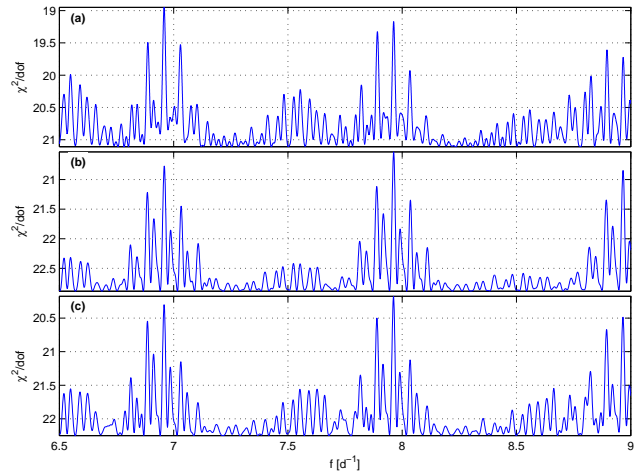}
\vskip 0.07in
\caption{{\bf (a)} The same as Fig.~\ref{Fig:V4633Sgr:PSyr} (g). {\bf
  (b)} A periodogram of an artificial LC of 2004,
  consisting of the fitted polinomial and signal of $P_1$, and
  randomly-selected residuals. {\bf (c)} A periodogram of an artificial
  LC of 2004, consisting of the fitted signals of $P_1$ and
  the 9.73 periodicity, a polynomial, and  randomly-selected
  residuals. The periodograms of the artificial sets were derived in
  the same way as the periodogram of the actual
  data.\label{Fig:V4633sgr:PS04Art}}
\end{figure}

The 2004 secondary periodogram shows no significant signal in the
expected frequency range of $P_2$.
The most significant frequency in the 6~d$^{-1}$-10~d$^{-1}$ is at
9.73~d$^{-1}$, the one-day alias of which at 8.73 is marked in
Fig.~\ref{Fig:V4633Sgr:PS1yr}.
Its $\frac{1}{2}$~day alias at 7.73~d$^{-1}$ is the most significant
peak in the 7.7~d$^{-1}$-7.9~d$^{-1}$ range, where $P_2$ is expected.
To confirm the absence of the secondary signal from the 2004 data, we
constructed an artificial LC of 2004 consisting of the
best-fitting primary signal and a polynomial, to which we randomly added
the residuals between the fitted and the observed data.
The periodogram of the artificial LC is very similar to that
of the observed one (Fig.~\ref{Fig:V4633sgr:PS04Art}, top and center
panels).
Another artificial data-set, that also comprises the fitted 9.73
signal yielded similar results (Fig.~\ref{Fig:V4633sgr:PS04Art}, bottom
panel).

\section{Discussion\label{Sec:V4633Sgr:Discussion}}
The analysis of 8-year data of V4633~Sgr allows us to review and refine
the main results presented in  \citetalias{LipkinEtAl2001}, as well as
to follow the continued post-eruption evolution of the nova.
We confirm the presence of two independent signals in the data, as
found by  \citetalias{LipkinEtAl2001}.
Both signals were present in the LC some 41 days after the eruption,
and possibly earlier.

\begin{figure}
\includegraphics[width=84mm]{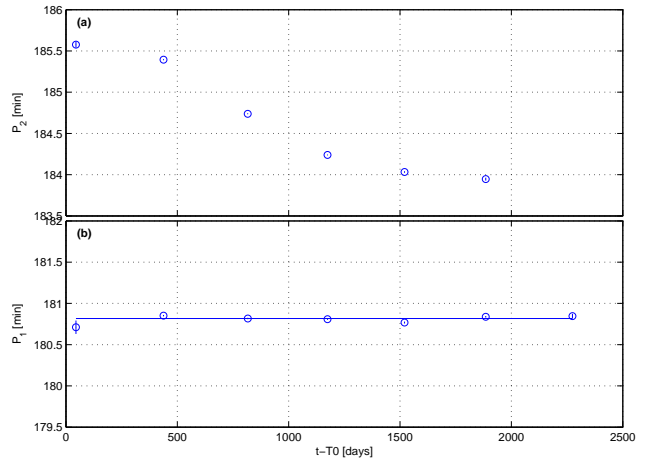}
\vskip 0.07in
\caption{The two signals in 1998--2004. The x-axis is the time since
  the nova eruption. The vertical lines crossing the circles are the
  error bars. The time evolution of $P_2$ is shown in the top
  panel. The bottom panel shows the measured yearly values of
  $P_1$. The solid line is the best-fitted value of $P_1$ in
  1999-2004\label{Fig:V4633Sgr:Periods}}
\end{figure}

The primary signal was detectable in all of the time-resolved
data-sets that followed, the latest of which was in 2004 August.
The period of the primary signal was confirmed to be stable over the
time base of our observations, (see Table~\ref{Table:V4633Sgr:Periods}
and Fig.~\ref{Fig:V4633Sgr:Periods}, lower panel), and coherent over the
5.3~yr time-base in which its coherence was tested for.
These characteristics support the interpretation of
\citetalias{LipkinEtAl2001} for this signal as the orbital period of
the underlying binary system.
The long time-base of the observations allowed us to refine the 
period of the primary signal, $P_1 = 180.8169\pm0.0002$~min.
The waveform of $P_1$, having an initial symmetric and then asymmetric
saw-tooth shape,had evolved into a symmetrical form with primary and
secondary lows.
These changes may reflect the changing geometry of the
binary (e.g., the contraction of the swollen primary), and
changes in the relative brightness of the system components.

The secondary signal, which was found to vary by
\citetalias{LipkinEtAl2001}, has decreased monotonically by 0.9\%
following the eruption, from being  2.6\% longer than $P_1$
some 40~days  after the eruption, to 1.7\% longer five years later.
(Table~\ref{Table:V4633Sgr:Periods} and Fig.~\ref{Fig:V4633Sgr:Periods},
upper panel).
In 1999-2003, the period decline-rate was also reduced monotonically by
an order of magnitude (from $\dot{P_2} \approx (-1.203 \pm 0.016)
\times 10^{-6}$ in 1999-2000 to $\dot{P_2} \approx (-1.7 \pm 0.4)
\times 10^{-7}$ in 2002-2003).
Finally, the secondary signal was undetectable in the data in
2004.

\citetalias{LipkinEtAl2001} proposed two possible interpretations to
the secondary signal: a permanent superhump, and more likely, the spin
period of a magnetic {\WD} of an asynchronous polar system.
Our results further weaken the permanent superhump interpretation.
The continued decline of $P_2$ further lowers the upper limit of the secondary
mass estimated by \citetalias{LipkinEtAl2001}, thus requiring an
exceptionally under-massed secondary, and an extremely
evolved CV system to be accounted for (\citetalias{LipkinEtAl2001}
Sec.~4.5 and references therein).
Moreover, the monotonic decrease of $P_2$ over a 5~yr time span is
atypical for a permanent superhump signal, which is expected to
wander randomly about a mean period that is determined by the
mass-ratio of the system.

Our results further support the AP model.
In particular, the post-eruption evolution of V4633~Sgr bares some
remarkable similarities to that of V1500~Cyg (Nova Cygni 1975) --
the  AP prototype, and one of the fastest and brightest novae of the
20th century ($t_2 = 2.43$~d, $m_{V,max} = 1.85$,
\citealt{YoungEtAl1976}).

Nova~Cygni displayed some unique photometric features following its
eruption.
A few days after outburst, a periodic variation, $\sim1$\%  longer
than the $P_{orb}=201$~min orbital period, was detected, initially in the
profiles of emission lines, and shortly after also photometrically
(\citealt{Patterson1978}; note that the photomeric orbital modulation
became detectably in photometry only 2 years after the eruption).
In the following 10 days or so, the period decreased by $\sim1$\%,
synchronizing with the orbital period.
The period continued to decrease by an additional $\sim1$\% during the
following year or so, after which the variation disappeared.
The periodicity, with a marginally shorter period, was rediscovered
some 10~yr later in the modulation of the optical circular
polarization \citep{Ref:V4633Sgr:Stockman_Schmidt_Lamb1988}, having an
increasing trend, with $\dot{P} =4.4\times10^{-8}$
\citep{Schmidt_Stockman1991}.

The polarized periodicity was naturally associated with the spin of a
strongly magnetic {\WD}, which in turn implied that the early
photometric variation was a tracer of the {\WD} spin evolution
following the nova eruption.
The commonly accepted scenario of this evolution was sketched by
\citet{Ref:V4633Sgr:Stockman_Schmidt_Lamb1988}.
They proposed that prior to the eruption, V1500~Cyg was an ordinary,
synchronized polar.
The rapid expansion of the strongly-coupled {\WD} envelope due to the
nova explosion increased the primary's moment of inertia, which lead,
owing to the conservation of angular momentum, to the {\WD} spin-down by
$\sim1$\%.
Strong magnetic coupling between the secondary and the expanded
envelope during the first few days after the eruption lead to a phase
of extremely rapid spin-up (with $\dot{P}\sim10^{-4}$), which resulted
in the resynchronization of the {\WD} spin and the orbital period.
The contraction of the remnant envelope onto the {\WD} surface over
the next year or so lead to a second spin-up phase, at a
rate that is slower by two orders of magnitude.
The photometric spin signal was switched off when the envelope finally
collapsed onto the {\WD} surface and residual nuclear burning
ceased.
The spin-down, with a synchronization time scale of $\sim150$~yr,
which was observed in polarimetry more than a decade later was
naturally interpreted as a magnetically-driven synchronization.

V4633~Sgr shares some striking observational similarities with
V1500~Cyg.
Shortly after their eruption, both novae exhibited a periodicity that was
slightly longer (by $\sim1$\% - 2.6\%)than the orbital period (in
V1500~Cyg within a week, and in V4633~Sgr after $\sim40$~d). 

In both systems the signal's period decreased monotonically at a rate
of $\dot{P}\sim10^{-6}$ that declined with time (the second spin-up
phase of V1500~Cyg). 
Finally, in both cases the unstable signal disappeared a short time
after the eruption ($\sim1.5$~yr and $\sim5.2$~yr, in V1500~Cyg and
V4633Sgr, respectively).

These similarities in the post-eruption evolution of V1500~Cyg and
V4633~Sgr further support the AP model that was proposed by
\citetalias{LipkinEtAl2001} for the latter.
Applying this model to V4633~Sgr, and assuming a post-eruption
evolution similar to that of V1500~Cyg, we note a number of differences
between the two systems.
The earliest measurement of the {\WD} spin of V4633~Sgr, some 40~days
after the eruption, yielded a spin period longer than $P_{orb}$ by
$\sim2.6$\% -- about twice the difference measured in V1500~Cyg
shortly after its eruption -- implying either a greater loss of
angular momentum in the eruption of V4633~Sgr, or a greater increase
of the {\WD} moment of inertia.
This would require either a stronger magnetic field, which would have
coupled the ejecta over a longer distance, and/or more mass
ejected during the eruption.
We note that because the brightness decline-rate of V4633~Sgr
($t_3\approx42$~d) was longer than in V1500~Cyg, the former's {\WD}
should be less massive \citepalias[Sec.~4.3 and references
therein]{LipkinEtAl2001}, and is therefore expected to eject more mass 
\citep{YaronEtAl2005}.

An extremely-rapid spin-up phase which was observed during the first
dozen days or so following the V1500~Cyg nova eruption was not
observed in V4633~Sgr, and particularly, no early resynchronization
has occurred.
As a result, the {\WD} spin period is longer than $P_{orb}$ in
V4633~Sgr, whereas in V1500~Cyg it is shorter.
The lack of early resynchronization implies that either the strong
coupling between the secondary and the envelope that was proposed by
\citet{Ref:V4633Sgr:Stockman_Schmidt_Lamb1988} for V1500~Cyg was
considerably weaker in V4633~Sgr.
Alternatively, if a rapid spin-up did occur, the desynchronization of
the {\WD} during the eruption of V4633~Sgr must have been
significantly stronger, implying a much greater loss of angular
momentum during the nova eruption.

The {\WD} spin period decreased over the 5~yr following the eruption by
0.9\% -- similar to the period decrease in the second spin-up phase of
V1500~Cyg.
If we naively assume an even-density contracting envelope surrounding a
{\WD}, a strong {\WD}-envelope coupling that effectively causes a
rigid-body behavior, angular-momentum conservation, and a
$10^{-5\,\mbox{-}\,-6} {{\rm M}_\odot}$ envelope, the observed
$\Delta{P_{rot}}$ would require an initial $(30 \mbox{-} 100)
{R_{\WD}}$ envelope -- a reasonable figure in the framework of the
\citet{Ref:V4633Sgr:Stockman_Schmidt_Lamb1988} model.

Finally, the envelope-contraction phase lasted $\sim5.28$~yr in
V4633~Sgr -- more than 4~times longer than in V1500~Cyg.
As pointed out by \citetalias{LipkinEtAl2001}, the longer
envelope-contraction time-scale also is expected for the less massive
{\WD} (\citealt{YaronEtAl2005}; note however that analysis of {\it
  XMM-Newton} observations by \citealt{Hernanz_Sala2007} suggests
that nuclear burning turned off less than 2.5~yr after the outburst).
 
The brightness modulation over the spin period suggests an intrinsic
brightness source on the {\WD} surface, which is non-uniformly
distributed.
The turning off in 2004 of this considerable source ($\sim25$\% of the
total brightness in 2003) may account for the renewed fading of the
system, perhaps because of reduced accretion due to decreased
irradiation of the secondary.
The source of the non-uniform brightness may be a hot spot due to
enhanced nuclear burning near the {\WD} accreting pole, as suggested
by \citet{Ref:V4633Sgr:Stockman_Schmidt_Lamb1988} for V1500~Cyg.
This would naturally account for the early appearance of the spin
modulation, less than 1.5~months after the eruption.
The disappearance of the signal 5~yr later may also be easily
explained by the final collapse of the envelope, and the termination
of the nuclear burning.

The contracting envelope -- the driving mechanism of the {\WD} spin up,
may also have contributed a significant fraction of the primary's
brightness variation.
Let us once again naively assume a {\WD} surrounded by an even-density
contracting envelope  with a mass $M_{env} = \alpha{\rm
  M}_{\WD}$ ($\alpha \approx 10^{-5\,\mbox{-}\,-6}$) 
and an initial post-eruption radius of $R_{env}  = \beta {\rm R}_{\WD}$
($\beta \approx 30 \mbox{-} 100$), a strong
{\WD}-envelope coupling leading to a rigid-body behavior, and
angular-momentum conservation.
The initial energy is the sum of the kinetic energy:
$$E_{K0} = \frac{1}{2}(I_{wd}+I_{env})\omega_0^2 \approx 
\frac{1}{2}I_{wd}\frac{\omega_f}{\omega_0}\omega_0^2 = \frac{1}{2}I_{wd}\omega_f\omega_0,$$
and the gravitational energy:
$$U_0 = -\int_{R_{wd}}^{R_{env}}{\frac{GM_{wd}M_{env}(r)}{r}dr} \approx$$
$$-2\pi GM_{wd}\frac{M_{env}}{\frac{4\pi}{3}(R_{env}^3-R_{wd}^3)}(R_{env}^2-R_{wd}^2) =$$
$$=-\frac{3}{2}GM_{wd}M_{env}\frac{R_{env}^2-R_{wd}^2}{R_{env}^3-R_{wd}^3} =
-\frac{3G M_{wd}^2}{2 R_{wd}}\alpha\frac{\beta^2-1}{\beta^3-1},$$
where $\omega_0$ and $\omega_f$ are the initial and final {\WD}
angular velocities, and $\frac{\omega_f-\omega_0}{\omega_0} \approx 2.6$\%.
The final energy is:
$$E_{Kf} + U_f \approx \frac{1}{2}I_{wd}\omega_f^2
-\frac{GM_{wd}M_{env}}{R_{wd}} = \frac{1}{2}I_{wd}\omega_f^2
-\alpha\frac{GM_{wd}^2}{R_{wd}},$$
and the energy released during the envelope contraction is:
$$\Delta E \approx \frac{1}{2}I_{wd}\omega_0^2(1-\frac{\omega_0}{\omega_f}) -
\alpha\frac{GM_{wd}^2}{R_{wd}}(1-\frac{3}{2}\frac{\beta^2-1}{\beta^3-1}) =$$
$$\frac{1}{5}M_{wd}R_{WD}^2\omega_0^2(1-\frac{\omega_0}{\omega_f}) -
\alpha\frac{GM_{wd}^2}{R_{wd}}(1-\frac{3}{2}\frac{\beta^2-1}{\beta^3-1}).$$
Inserting typical {\WD} mass and radius, $\alpha$ and $\beta$, we obtain:
$$\Delta E \approx 1.4\cdot10^{40}~erg -
\alpha(1-\frac{3}{2}\frac{\beta^2-1}{\beta^3-1})3.8\cdot10^{50}~erg \approx$$
$$\approx -10^{44-45}erg$$

These order of magnitude calculations show that only a marginal
fraction of the gravitational energy release that is associated with
the contraction of the envelope is required for the spinning up of the
{\WD}.
Almost all of it, an amount of energy sufficient to maintain an
Eddington luminosity of $10^{38}$~erg~s$^{-1}$ for for 0.1-1~years, is
radiated away during the envelope's contraction.
Because of the strong magnetic field of the {\WD} that couples it, the
hot, ionized envelope cannot contract radially, but is forced to
accrete along the magnetic-field lines towards the magnetic poles of
the {\WD}.
It is thus reasonable that most of this energy is released near the
magnetic poles, becoming a major energy source for a hot spot near one
or the two poles, and a significant fraction of the variable brightness
component on the white dwarf (and indeed, of total luminosity of
the system) during these first 5 years.
The heat released at the impinge of the falling materials may also
serve as a catalyst to intensify the residual nuclear burning of the
hydrogen-rich material that accrets at these locations.
The gravitational energy release near the poles, as well as the
intense nuclear burning there, are expected to begin shortly after the
eruption, when the envelope starts contracting.
They also should terminate together, when the envelope finally
collapses onto the {\WD} surface.

\section{Summary\label{Sec:V4633Sgr:Summary}}
\begin{enumerate}

\item Time resolved photometry of V4633~Sgr between 1998 and 2004
  (three months to six years following the nova eruption) reveals two
  ptotometric periodicities in the LC.

\item The shorter periodicity, $P_1 = 180.8$~min was stable, and is
  identified as the orbital period of the underlying binary system.

\item The second periodicity has been shortened monotonically, from
  $P_2 = 185.6$~min a few weeks after the eruption to $P_2 = 183.9$~min five
  years later.
  We identify this periodicity with the spin of the {\WD} primary in a
  magnetic CV, and the decrease as the manifestation of the {\WD}
  spin-up due to the contraction of its envelope.

\item In 2004 the latter period was no longer detectable in the LC.
  In the near-synchronous scenario, this marks the final decline of
  the {\WD} envelope and the termination of residual nuclear burning on
  the {\WD} surface.

\item V4633 Sgr thus seems to be the second known asynchronous polar classical nova.
\end{enumerate}

\section*{ACKNOWLEDGMENTS\label{Sec:V4633Sgr:Acknowlegments}}
We are grateful to Dina Prialnik and Ofer Yaron for some very useful
discussions.
This work has been supported by the Israel Science Foundation.

\label{lastpage}
\end{document}